\documentclass{IEEEtran4PSCC}
\ifCLASSINFOpdf
   \usepackage[pdftex]{graphicx}
\else
   \usepackage[dvips]{graphicx}
\fi
\usepackage[cmex10]{amsmath}
\usepackage{color,soul}

\usepackage[utf8]{inputenc}
\usepackage{cite}
\usepackage[usenames,dvipsnames]{xcolor}
\usepackage[pdftex]{graphicx}
\usepackage{circuitikz}

\usepackage{booktabs,colortbl}
\usepackage{multirow}
\usepackage[cmex10]{amsmath}
\usepackage{mathrsfs}
\usepackage{mathtools}

\usepackage{array}

\usepackage{siunitx}
\usepackage{amssymb}
\usepackage{float}

\usepackage{algpseudocode}
\usepackage{url}
\usepackage{multirow}
\usepackage{color,soul}
\usepackage{url}
\usepackage{setspace}

\newcommand{\quantities}[1]{%
  \begin{tabular}{@{}c@{}}\strut#1\strut\end{tabular}%
}

\usepackage[ruled, vlined]{algorithm2e}

\makeatletter
\let\old@ps@headings\ps@headings
\let\old@ps@IEEEtitlepagestyle\ps@IEEEtitlepagestyle
\def\psccfooter#1{%
    \def\ps@headings{%
        \old@ps@headings%
        \def\@oddfoot{\strut\hfill#1\hfill\strut}%
        \def\@evenfoot{\strut\hfill#1\hfill\strut}%
    }%
    \def\ps@IEEEtitlepagestyle{%
        \old@ps@IEEEtitlepagestyle%
        \def\@oddfoot{\strut\hfill#1\hfill\strut}%
        \def\@evenfoot{\strut\hfill#1\hfill\strut}%
    }%
    \ps@headings%
}
\makeatother

\psccfooter{%
        \parbox{\textwidth}{\hrulefill \\ \small{21st Power Systems Computation Conference} \hfill \begin{minipage}{0.2\textwidth}\centering \vspace*{4pt} \includegraphics[scale=0.06]{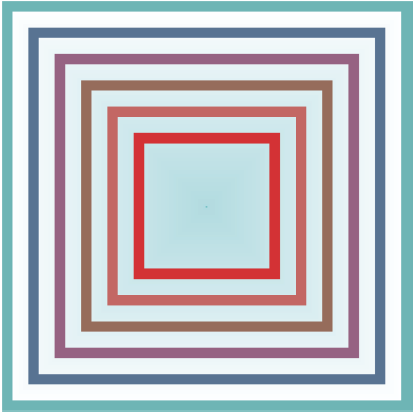}\\\small{PSCC 2020} \end{minipage} \hfill \small{Porto, Portugal --- June 29 -- July 3, 2020}}%
}

\begin{document}
%
\title{Optimal Provision of Concurrent Primary Frequency and Local Voltage Control from a BESS Considering Variable Capability Curves: Modelling and Experimental Assessment}

\author{
\IEEEauthorblockN{Antonio Zecchino, Zhao Yuan, \\ Rachid Cherkaoui, Mario Paolone}
\IEEEauthorblockA{Distributed Electrical Systems Laboratory \\
EPFL - Lausanne, Switzerland\\
\{antonio.zecchino, zhao.yuan\}@epfl.ch}
\and
\IEEEauthorblockN{Fabrizio Sossan}
\IEEEauthorblockA{Centre for processes, renewable energies and energy systems \\
MINES ParisTech - Nice, France}
}


\maketitle

\begin{abstract}
This paper proposes a control method for battery energy storage systems (BESSs) to provide concurrent primary frequency and local voltage regulation services.
The actual variable active and reactive power capability of the converter, along with the state-of-charge of the BESS, are jointly considered by the optimal operating point calculation process within the real-time  operation. 
The controller optimizes the provision of grid services, considering the measured grid and battery statuses and predicting the battery DC voltage as a function of the current trajectory using a three-time-constant model (TTC). A computationally-efficient algorithm is proposed to solve the formulated optimal control problem.
Experimental tests validate the proposed concepts and show the effectiveness of the employed control framework on a commercial utility-scale 720 kVA/560 kWh BESS.
\end{abstract}


\thanksto{\noindent This work was supported by the European Union’s
Horizon 2020 research and innovation program under agreement no. 773406.}

\section{Introduction}
Battery energy storage systems (BESSs) are broadly recognized as essential assets for the operation of modern power systems thanks to their wide controllability and power ramping rate that can be exploited for grid balancing regulation purposes~\cite{Hornsdale_Report}.
As extensively demonstrated in the literature, one of the most popular power system services achieved by BESS is primary frequency control (PFC), which is increasingly needed from transmission system operators (TSOs) given the progressive displacement of conventional generation plants in favor of stochastic renewable-based generation units~\cite{Ach_dis, coordinated_BESS_ctrl, BESS_in_freq_ctrl_market, Sandgani, Gundogdu}. 
PFC is typically performed by a frequency droop controller that determines the variation of the active power $(\Delta P)$ exchanged with the AC grid for a given frequency deviation from a reference value. 
Since power converters are normally able to operate on the 4 quadrants of their PQ capability curve, they are also capable of exchanging reactive power concurrently with the active power.
Within this context, the proposed control approach considers the additional simultaneous exchange of reactive power, which is seen as a viable mean for local voltage regulation at distribution grid level~\cite{BESS_PQ_control, Serban, Opt_PV, Vctrl_PV_OLTC, ZECCHINO_Q_effects}. 
Similarly to the case of PFC, local voltage deviations from the nominal value can be used as input for determining the necessary variation on reactive power $(\Delta Q)$. 
Although a dedicated market framework is not existing at distribution grid level, distribution system operators (DSOs) are yet keen to have the possibility of acquiring voltage regulation services via reactive power support coming from distributed energy resources, e.g.,  by imposing relevant requirements to photovoltaics plants~\cite{Ger_PV_codes, It_PV_codes}.

The proposed joint PFC-voltage control actions are achieved within the real physical constraint of having a non-unique PQ region of feasibility of the BESS power converter: this region is in fact a function of the battery DC-link and AC grid statuses. 
This aspect goes beyond the typical assumptions present in the existing scientific literature where it is assumed that the PQ capability curve of the BESS converter is static and does not depend on battery state-of-charge (SOC) and AC grid voltage conditions.
In fact, to the best of the Authors' knowledge, neither the studies that consider only active power control (e.g.,~\cite{Ach_dis, coordinated_BESS_ctrl, BESS_in_freq_ctrl_market, Sandgani, Gundogdu}) nor those including concurrent reactive power control (e.g.,~\cite{BESS_PQ_control, Serban}) do take into account limitations given by the variability of the BESS converter capability curve.
Another contribution proposed in this work is the design of a real-time control algorithm with a time execution $\leq1 s$, capable of computing the maximum grid support from the BESS as function of the battery and AC grid conditions. 
This is done by solving, in real time, an optimization problem that includes constraints based on battery status predictions, given by a three-time-constant (TTC) equivalent model, whose parameters have been experimentally estimated via a series of model identification tests \cite{Vdc_estimation, Emil_ISGT}. 
In this approach, the nonconvex constraints are relaxed and convexified to make them easier to be solved. Similar approaches have been also used and validated to solve power system operation problems in \cite{DLMP_me, SOCACOPF_relax_fea_me}.
Ultimately, the paper reports an experimental validation of the proposed control approach. 
Finally, the real-time optimal controller is implemented and tested on the utility-scale 720 kVA/560 kWh BESS installed at EPFL campus in Lausanne, Switzerland.

In sum, the main research contributions of this paper are three-fold: \textit{i)} propose a method that accounts for the variability of the feasibility PQ region of the BESS power converter as function of both the AC grid and internal BESS conditions; \textit{ii)} develop a control framework for concurrent provision of power system frequency and local voltage control based on the real-time solution of an optimization problem to maximize the contribution to grid support that the BESS can provide for given actual and predicted operating conditions; \textit{iii)} provide the experimental validation of the proposed method in a real grid.  
The paper is structured as follows: Section II outlines the proposed methodology. In Section III, the utility-scale BESS deployed for the validation experimental activities is described, including the estimations of the identified parameters of the equivalent battery model. Section IV reports the results of relevant experimental test study cases. Conclusions and possible future works are included in Section VI.

\section{Proposed Methodology}

The BESS converter is controlled to provide primary frequency and local voltage regulation adjusting the active and reactive power set-points, respectively. The initial power set-points are achieved via droop logics:
\begin{align}
    P^{AC}_{0,t}= \alpha_0 \Delta f_{t};\; Q^{AC}_{0,t}= \beta_0 \Delta v^{AC}_{t} , \label{eq:droop_p_q}
\end{align}
where $t \in T$ is the discrete index of time, $P^{AC}_{0,t}, Q^{AC}_{0,t}$ are the initial active and reactive power set-points that the BESS will set for given grid frequency and AC voltage magnitude deviations from their nominal values $(\Delta f_{t}, \Delta v^{AC}_{t})$, according to the initial droop coefficients $\alpha_0, \beta_0$. These active and reactive power set-points will be adjusted when considering the converter capability curves, as described later in the paper.

To maximize the frequency and voltage regulation performance, the initial droop coefficients $\alpha_0, \beta_0$ can be set as:
\begin{align}
    \alpha_0 =\frac{P^{max}}{\Delta^{max} f_{t}};\; \beta_0 = \frac{Q^{max}}{\Delta^{max} v^{AC}_{t}} ,\label{eq:droop_set}
\end{align}
where $P^{max}$ and $Q^{max}$ are the maximum active and reactive power that the BESS can exchange, as specified by the BESS technical specifications. 
Historical measurements can be used to determine the maximum frequency and voltage deviation $\Delta^{max} f_{t}, \Delta^{max} v^{AC}_{t}$, as shown in Section III. During real-time operations, the employed $\alpha_t, \beta_t$ are adjusted by relying on BESS status (available storage capacity and SOC) and solving an optimal power set-points calculation problem \eqref{eq:opt_pq}.

\begin{figure}[!ht]
		\centering
    	\includegraphics[width=0.9\linewidth]{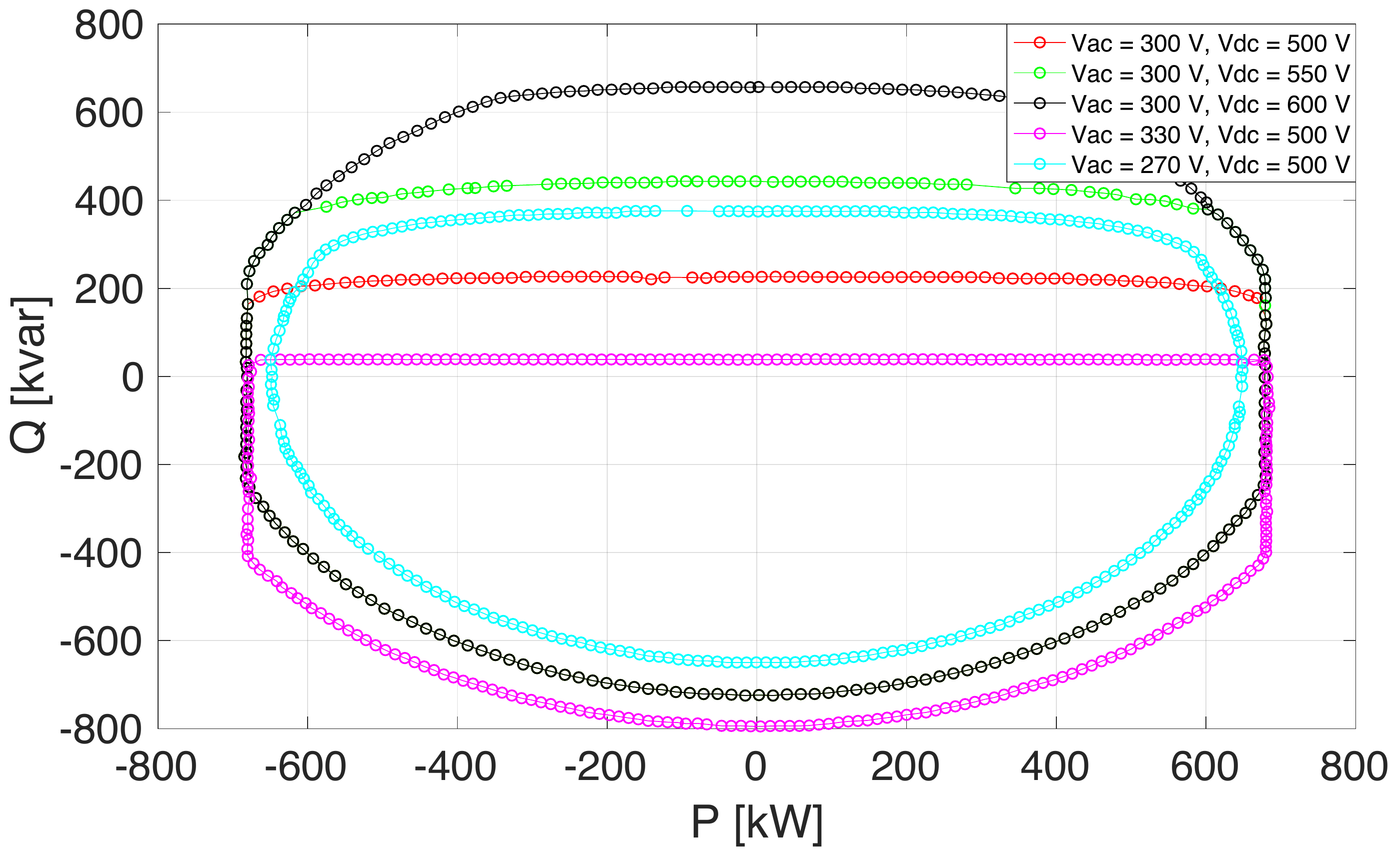}
		\caption{BESS converter PQ capability curves as function of $v^{AC}_t$ and $v^{DC}_t$.}
		\label{fig:Converter_CapabilityCurves}
\end{figure}

Commonly, in the current literature the converter capability is considered to be constantly expressed as $(P^{AC}_t)^2+(Q^{AC}_t)^2 \leq (S^{AC})^2$, where $P^{AC}_t, Q^{AC}_t$, and $S^{AC}$ are the converter output active, reactive and maximum apparent power of the grid converter, respectively. This assumption, however, does not hold in practice.
In this work, the realistic feasible operation region identified by the PQ converter capability curves $h$ in Fig. \ref{fig:Converter_CapabilityCurves}, are considered as: 
\begin{align}
    h(P^{AC}_t, Q^{AC}_t, v^{DC}_t, v^{AC}_t, SOC_{t}) \leq 0 \label{eq:bound_pq}
\end{align}
being $v^{DC}_t$ the voltage of the BESS DC bus and $v^{AC}_t$ the module of the direct sequence component of the phase-to-phase voltages at the AC side. 
Notably, the capability curves $h$ are specific for the employed hardware, but similar dependencies are expected in all kinds of utility-scale BESS converters. 
More detailed information about the PQ curves considered in this study are included in Section III.

\vspace{-0.3 cm}

\vspace{-0.1 cm}

\begin{figure}[ht]
		\centering
    	\includegraphics[width=0.6\linewidth]{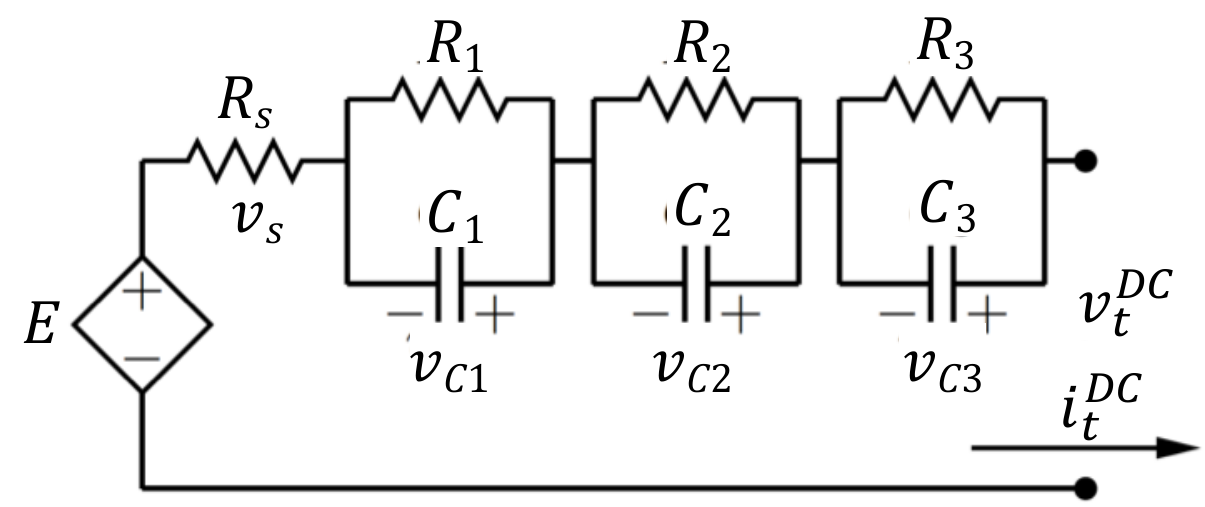}
		\caption{Three time constant TTC model.}
		\label{fig:TTC_model}
\end{figure}

The $v^{DC}_t$ voltage needed for the selection of the capability curve is estimated via the TTC model shown in Fig. \ref{fig:TTC_model}, whose parameters are derived by dedicated model identification tests.
Since the BESS has to be controlled in a very small time resolution ($\leq1 s$), we estimate the battery status based on the TTC model state equations:
\begin{align}
    &C_1\frac{\mathrm{d}v_{C1} }{\mathrm{d} t}+\frac{v_{C1}}{R_1}=\frac{v_s}{R_s} \label{eq:ttcvc1}\\
    &C_2\frac{\mathrm{d}v_{C2} }{\mathrm{d} t}+\frac{v_{C2}}{R_2}=\frac{v_s}{R_s}\label{eq:ttcvc2}\\
    &C_3\frac{\mathrm{d}v_{C3} }{\mathrm{d} t}+\frac{v_{C3}}{R_3}=\frac{v_s}{R_s}\label{eq:ttcvc3}\\
    &v_s+v_{C1}+v_{C2}+v_{C3}=E-v^{DC}_{t} , \label{eq:ttcsum}
\end{align}
where $\mathbf{v_c}=[v_{C1};\, v_{C2};\, v_{C3}]$ are the TTC state voltage variables that are updated by solving \eqref{eq:ttcvc1}-\eqref{eq:ttcsum} in each control loop. At each time step, the initial value of the state variables can be estimated via the use of dedicated state observers as proposed in~\cite{Ach_dis}. The model \eqref{eq:ttcvc1}-\eqref{eq:ttcsum} is discretized at a 1s resolution in this paper.
The TTC model capacitance parameters $C_1, C_2, C_3$ and resistance parameters $R_s, R_1, R_2, R_3$ are identified by generating active power pseudo random binary signals (PRBS) and then by measuring the corresponding current dynamics. 
This process is explained in Section III. 
The voltage source $E$ is the open circuit voltage of the battery, which depends on the SOC as shown in \eqref{eq:E_SOC}. 
$E$ is modelled as a linear function of the battery SOC, where the parameters $a$ and $b$ are identified within the TTC model identification process. 
\begin{align}
E(SOC_t)&=a + b\cdot SOC_t
\label{eq:E_SOC}  
\end{align}

After updating $\mathbf{v_c}=[v_{C1};\, v_{C2};\, v_{C3}]$, considering $v_s=\frac{P^{DC}_t}{v^{DC}_t} R_s$, equation \eqref{eq:ttcsum} is equivalent to:
\begin{align}
    (v^{DC}_t)^2+(\mathbf{1}^T\mathbf{v_c}-E)v^{DC}_t+P^{DC}_t R_s=0 , \label{eq:vdcpdc}
\end{align}
where $\mathbf{1}^T=[1, 1, 1]$. Solving constraint \eqref{eq:bound_pq} jointly with \eqref{eq:vdcpdc} gives feasible power set-points $P^{AC}_{t},Q^{AC}_{t}$ satisfying the evolving capability curves during the control loop.

Given the initial state-of-charge $SOC_{0}$, its value at each discrete time control iteration, $SOC_{t}$, can be expressed as:
\begin{align}
SOC_t&=SOC_{t-1}+ \frac{\int_{t-1}^{t}i^{DC}_t dt}{C^{max}} \nonumber\\
&\approx SOC_{t-1}+ \frac{P^{DC}_t}{v^{DC}_{t} C^{max}} \Delta t , \label{eq:socupd}  
\end{align}
where $C^{max}$ is the maximum storage capacity of the battery in Ampere-per-hour and $i^{DC}\approx \frac{P^{DC}_t}{v^{DC}_{t}}$ is the charging or discharging DC current. The active power at the DC bus $P^{DC}_t$ is related the active power at the AC side of the converter as:
\begin{align}\label{eq:pdcac}
P^{DC}_t=\left\{\begin{matrix}
\eta P^{AC}_t ,\; \forall P^{AC}_t<0\\ 
\frac{P^{AC}_t}{\eta},\; \forall P^{AC}_t\geq0 
\end{matrix}\right. , 
\end{align}
where $\eta=97\%$ is the efficiency of converter. $P^{AC}_t<0$ means charging of the BESS and $P^{AC}_t\geq0$ means discharging. The state-of-charge $SOC_t$ should be always kept in the secure limits during all the operational periods $t \in T$:
\begin{align}
    SOC^{min}\leq SOC_t \leq SOC^{max} \label{eq:bound_soe}
\end{align}

\begin{figure}[ht]
		\centering
    	\includegraphics[width=0.7\linewidth]{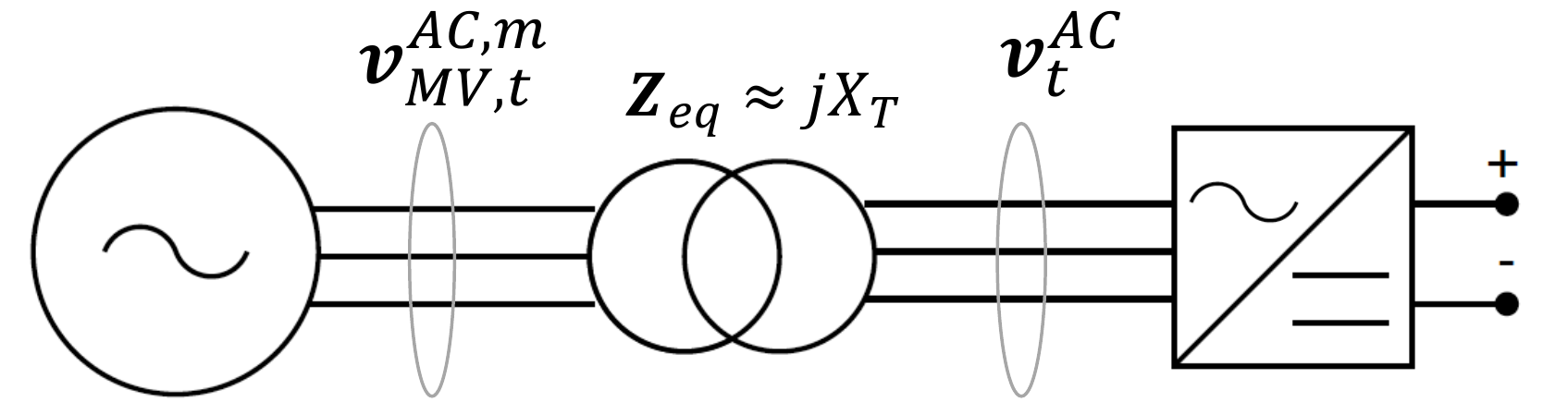}
		\caption{Reference BESS scheme for the AC voltage prediction.}
		\label{fig:Thev_eq}
\end{figure}

The magnitude of the direct sequence component $v^{AC}_t$ of the phase-to-phase voltages needed for the selection of the converter capability curve is estimated via the Thévenin equivalent circuit of the AC grid.
As shown by Equation \eqref{eq:thev_eq}, the estimation considers the direct sequence component $\mathbf{v}^{AC,m}_t$ of the measured phase-to-phase voltages and the expected voltage drop due to the three-phase complex power $\mathbf{S}^{AC}_{0,t}$ exchanged by the BESS over the grid equivalent impedance $\mathbf{Z}_{eq}$. $\mathbf{Z}_{eq}$ can be approximated as the BESS step-up transformer reactance $jX_T$.
Since, as shown in Fig. \ref{fig:Thev_eq}, measurements are acquired at the primary side of the BESS step-up transformer whereas the estimation is done for the voltage at the secondary side, the voltage $\mathbf{v}^{AC,m}_t$ in \eqref{eq:thev_eq} is referred to the secondary side as $\mathbf{v}^{AC,m}_t=\mathbf{v}^{AC,m}_{MV,t}\frac{1}{n}$, being $n$ the transformer ratio. 

\vspace{-0.1 cm}

\vspace{-0.1 cm}

\begin{align}
\mathbf{v}^{AC}_t=\mathbf{v}^{AC,m}_t+ \mathbf{Z}_{eq} \textit{conj}(\frac{\mathbf{S}^{AC}_{0,t}}{\sqrt{3}\mathbf{v}^{AC,m}_t}) \nonumber\\
v^{AC}_t \approx \sqrt{(v^{AC,m}_t)^2+ X^{2}_{T}\frac{(P^{AC}_{0,t})^2+(Q^{AC}_{0,t})^2}{3(v^{AC,m}_t)^2}} 
\label{eq:thev_eq}  
\end{align}

The optimal active and reactive power set-points are given by solving the following optimization problem: 
\begin{align}
    &\text{Minimize}\; \lambda_{P}( P^{AC}_{t}- P^{AC}_{0,t})^2+\lambda_{Q}( Q^{AC}_{t}- Q^{AC}_{0,t})^2 \label{eq:opt_pq}\\
    &\text{subject to}\;\; \eqref{eq:droop_p_q} -\eqref{eq:bound_pq}, \eqref{eq:vdcpdc}-\eqref{eq:thev_eq}\nonumber
\end{align}
Where $\lambda_{P}$ and $\lambda_{Q}$ are weight coefficients used by the modeler to prioritize the provision of active or reactive power, i.e., to prioritize one grid service over the other.
In the case of equal priority for frequency and voltage control, the weight of 1 is assigned to both coefficients, meaning that the optimal power set-points $ P^{AC}_{t}, Q^{AC}_{t}$ are the closest to the initial power set-points $P^{AC}_{0,t}, Q^{AC}_{0,t}$ inside the feasible operational region of the BESS defined by \eqref{eq:droop_p_q}-\eqref{eq:bound_pq} and \eqref{eq:vdcpdc}-\eqref{eq:bound_soe}. After finding the optimal power set-points $P^{*AC}_t, Q^{*AC}_t$ , the optimal droop parameters $\alpha^{*}_{t},\beta^{*}_{t}$ are defined as:
\begin{align}
    \alpha^{*}_{t} =\frac{P^{*AC}}{\Delta f_{t}};\; \beta^{*}_{t} = \frac{Q^{*AC}}{\Delta v^{AC}_{t}}\label{eq:droop_opt}
\end{align}

This optimization problem is nonconvex due to the nonconvex constraints \eqref{eq:vdcpdc}, \eqref{eq:socupd} and \eqref{eq:pdcac}. 
To efficiently find a local optimal solution, constraint \eqref{eq:vdcpdc} is firstly convexified to:
\begin{align}
    (v^{DC}_t)^2+(\mathbf{1}^T\mathbf{v_c}-E)v^{DC}_t+P^{DC}_t R_s\leq0 \label{eq:vdcpdc2}
\end{align}
This relaxation shows better computational efficiency in real-time control experiments. 
Then, to find the optimal power set-points, we propose the computationally-efficient solution algorithm shown in Algorithm 1, where $V^{DC}_i\in \left \{(500, 550], (550, 600], (600, 800]\right \}$ is $i$-th set of the DC voltage range and where $V^{AC}_j\in \left \{(270, 330], (330,+Inf)\right \}$ is $j$-th set of the AC voltage range. 
Algorithm \ref{alg:heuris} works by firstly assuming the ranges that could include the DC voltage $v^{DC}_t$ and the AC voltage $v^{AC}_t$ solutions. 
Then, one capability curve is selected based on the assumed DC voltage and the predicted AC voltage. 
If the calculated $v^{DC}_t$ and $v^{AC}_t$ are consistent with the initial assumed DC and AC voltage ranges $V^{DC}_i$ and $V^{AC}_j$, the algorithm converges.
Otherwise, the assumption of the DC and AC voltage ranges is changed and another capability curve is selected until a consistent solution is found.

\begin{algorithm}\label{alg:heuris}
\setstretch{1}
 \caption{Optimization Solution Algorithm}
\SetAlgoLined
\KwResult{Optimal Power Set-Points $P^{AC}_{t}, Q^{AC}_{t}$}
 Initialization $P^{AC}_0, Q^{AC}_0, i=1, j=1$\;
 \While{$v^{DC}_t \notin {V^{DC}_i}$ and $i<i^{max}$}{
  Assume $v^{DC}_t \in V^{DC}_i$\;
  \While{$v^{AC}_t \notin {V^{AC}_j}$ and $j<j^{max}$}{
  Assume $v^{AC}_t\in V^{AC}_j$\;
  Select one capability curve:
  $h(P^{AC}_t, Q^{AC}_t, v^{DC}_t, v^{AC}_t, SOC_t)$\;
  \eIf{$P^{AC}_0<0$}{
   $P^{DC}_t=\eta P^{AC}_t$\;
   }{
   $P^{DC}_t=\frac{P^{AC}_t}{\eta}$\;
  }
  \text{Solve the Optimization Problem}\;
  $j=j+1$\;
  }
  $i=i+1$\;
  }
\end{algorithm}

\begin{figure}[ht]
		\centering
    	\includegraphics[width=1\linewidth]{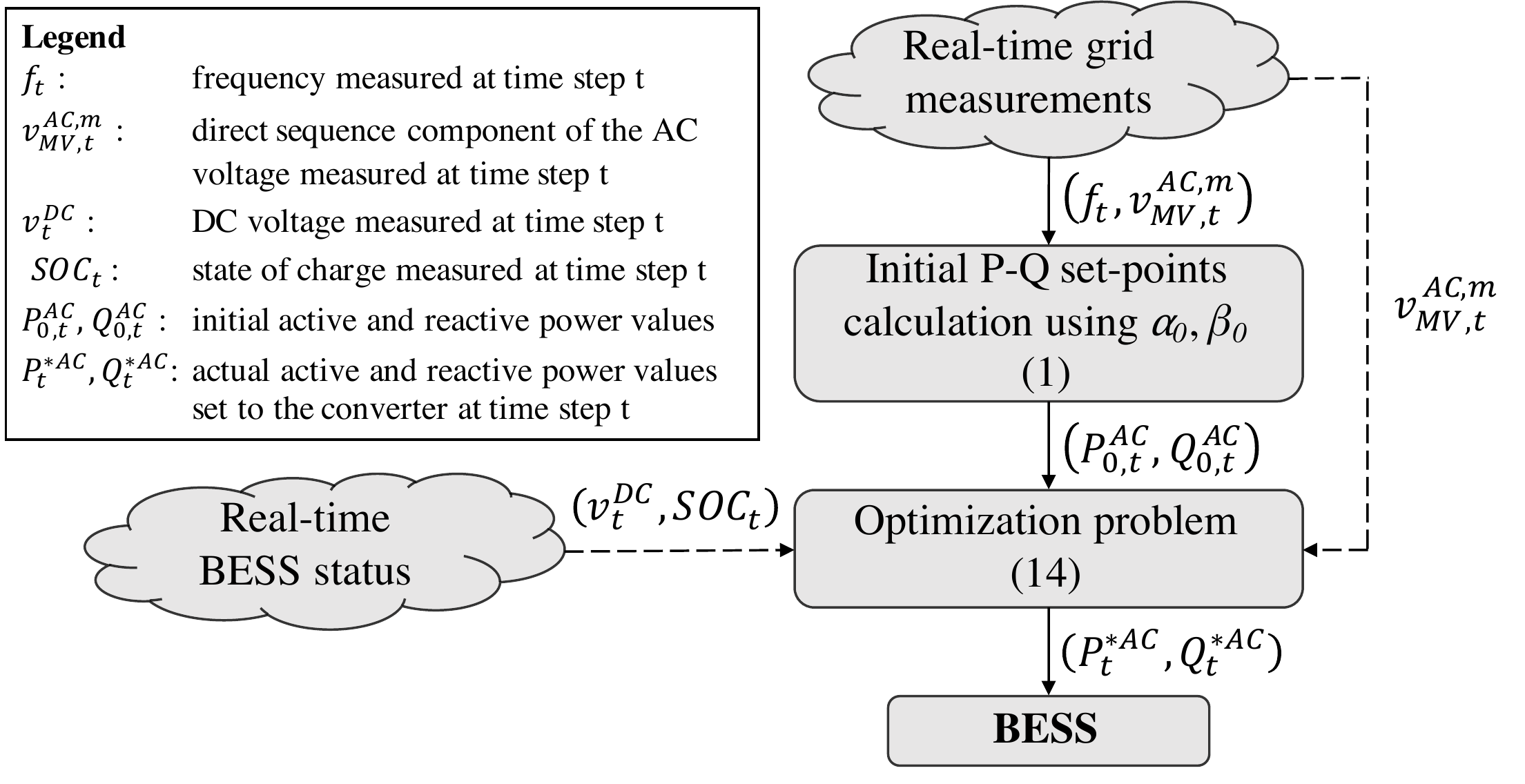}
		\caption{Block diagram of the proposed real-time controller.}
		\label{fig:ctrl_framework}
\end{figure}

The block diagram of the proposed controller during one time step is illustrated in Fig. \ref{fig:ctrl_framework}.

\section{Utility-Scale BESS Capability Curves and Assessment of its Equivalent Circuit Model}
The testbed of the proposed validation study consists in an utility-scale BESS installed at the EPFL campus in Lausanne, Switzerland. The system is based on a 720 kVA/560 kWh Lithium-Titanate-Oxide (LTO) battery, utilized for a number of power grid support experimental activities~\cite{Namor_thesis}.
The BESS is equipped with a 720 kVA 4-quadrant converter, which can be controlled via \textit{Modbus TCP} with a refresh rate up to 50 ms.
The BESS is connected to one of the feeders of the EPFL campus medium voltage (MV) grid via a 630 kVA 3-phase 0.3/21 kV step-up transformer. 
The paramenters of the main components of the employed BESS are reported in Table~\ref{BESS}.
The selected MV feeder presents all the peculiarities of modern active distribution grids: the lines are relatively short, the load demand is largely variable during the day (office buildings with 300 kWp), and a substantial amount of rooftop PV units is connected (for a total of 95 kWp). 
Such characteristics make the testbed suitable for investigations not only on system frequency regulation, but also on local voltage control solutions such as the one proposed in this work.

\begin{table}[!ht]
\caption{Specifications of the employed utility-scale BESS}
\label{BESS}
\centering
\renewcommand{\arraystretch}{1}
\begin{tabular}{|c|c|} \hline
Parameter                    & Value      \\ \hline
Energy Capacity               &560 kWh           \\
Maximum Power                 &720 kVA            \\
Nominal Active Power          &640 kW            \\
Rated AC grid voltage            &0.3 kV, three-phase            \\
Maximum AC current            &1385 A            \\
AC current distortion (THD)   &3$\%$            \\
Nominal DC voltage            &750 V            \\
DC voltage range              &500-890 V            \\
Inverter efficiency                    &$\geq$97$\%$            \\
Transformer rated power                   &630 kVA            \\
Transformer high voltage                  &3 x 21 kV            \\
Transformer low voltage                   &3 x 0.3 kV            \\
Transformer short-circuit voltage       &6.28$\%$            \\
Transformer group                         &Dd0            \\\hline
\end{tabular}
\end{table}

As commonly known, a peculiarity of BESS installations is the modular structure. 
For the specific commercial MW-class BESS under analysis, 3 series of 20 cell elements are connected in parallel to compose the battery module, 15 modules in series compose one string, and finally 9 strings connected in parallel guarantee the desired BESS energy storage and power capacity. 
The main advantage of such modular structure is the absence of the limiting single point of failure typical of conventional power grid service providers.
In fact, the system can be operated even if one element is not correctly in operation. 
Within this context, in the analysis proposed in this work a configuration with reduced number of strings is considered. Specifically, 7 strings out of 9 are utilized, meaning that the available storage capacity is 7/9 of the total value, i.e., 435 kWh. 
One has to note that the reduced number of usable strings should be considered also in the setting of the maximum power exchange capability, being the strings connected in parallel.
This is done to prevent string over-currents and over-temperatures, without jeopardizing the cycle aging process of the cells.
In this respect, at the implementation stage of the controller, the constraints of the power converter PQ capability curves presented in Fig. \ref{fig:Converter_CapabilityCurves} have been shrank by the factor $C_{shrink}$, which in this case is 7/9.

As shown in Fig. \ref{fig:Converter_CapabilityCurves}, the region of feasible operating points of the power converter depends on the grid AC voltage and on the DC battery voltage in a non-linear way. 
In fact, for increasing battery DC voltages only the maximum positive Q value is increasing. 
The curve is shifted down vertically for AC voltages higher than the nominal value, meaning that both the maximum positive Q is decreased, whereas the maximum negative Q is increased.
A different pattern is present for AC voltages lower than the nominal value: the limit values are shrank both for the active and the reactive part of the apparent power set-point in both negative and positive signs.
At the implementation stage of the proposed controller, the dependency of the feasibility region on the grid and battery statuses is considered in a discretized way, by selecting two of the five PQ curves and by considering the overlapping area between them. 
As mentioned, this is done in accordance with the respective factor $C_{shrink}$. The capability curves of the employed power converter are fitted using datasheet information from the manufacturer and, then, scaled proportionally to the available BESS capacity. 
The fitted capability curves consist of a series of linear and quadratic functions, which are reported in Table~\ref{tab:Coeff}.

\begin{table}[!ht]
\caption{Fitted Functions of the Converter PQ Capability Curves}
\centering
\renewcommand{\arraystretch}{1}
\begin{tabular}{|c|c|c|}\hline
 $v^{DC}$ &$v^{AC}$  &Functions    \\\hline
    \multirow{6}[0]{*}{600 V} & \multirow{6}[0]{*}{300 V} & $P\geq-681.89$  \\
          &       & $P\leq678.71$   \\
          &       & $P^2+Q^2 \leq723.03^2,\forall Q\geq0$   \\
          &       & $P^2+Q^2 \leq719.19^2,\forall Q<0$  \\
          &       & $Q\leq 659.67-8.29^{-18} P -2.16^{-4} P^2$   \\
          &       & $Q\leq 657.1$  \\\hline
    \multirow{6}[0]{*}{550 V} & \multirow{6}[0]{*}{300 V} & $P\geq-681.89$  \\
          &       & $P\leq678.71$  \\
          &       & $P^2+Q^2 \leq723.03^2,\forall Q\geq0$  \\
          &       & $P^2+Q^2 \leq717.93^2,\forall Q<0$  \\
          &       & $Q\leq 459.43 -1.5^{-3} P -2.12^{-4} P^2$ \\
          &       & $Q\leq439.98$  \\\hline
    \multirow{5}[0]{*}{500 V} & \multirow{5}[0]{*}{300 V} & $P\geq-680.62$  \\
          &       & $P\leq682.45$   \\
          &       & $P^2+Q^2 \leq721.4^2 $  \\
          &       & $Q\leq 286.64 + 1.4^{-3} P + -2.33^{-4} P^2$ \\
          &       & $Q\leq225.22$   \\\hline
    \multirow{4}[0]{*}{500 V} & \multirow{4}[0]{*}{330 V} & $P\geq-679.21$  \\
          &       & $P\leq681.06$ \\
          &       & $P^2+Q^2 \leq794.34^2$   \\
          &       & $Q\leq38.47 $ \\\hline
    \multirow{2}[0]{*}{500 V} & \multirow{2}[0]{*}{270 V} & $P^2+Q^2 \leq649.5^2 $  \\
          &       & $Q\leq382.95 + 1.6^{-3} P -2.21^{-4} P^2$\\\hline
\end{tabular}\label{tab:Coeff}
\end{table}

\begin{figure}[ht]
		\centering
    	\includegraphics[width=0.90\linewidth]{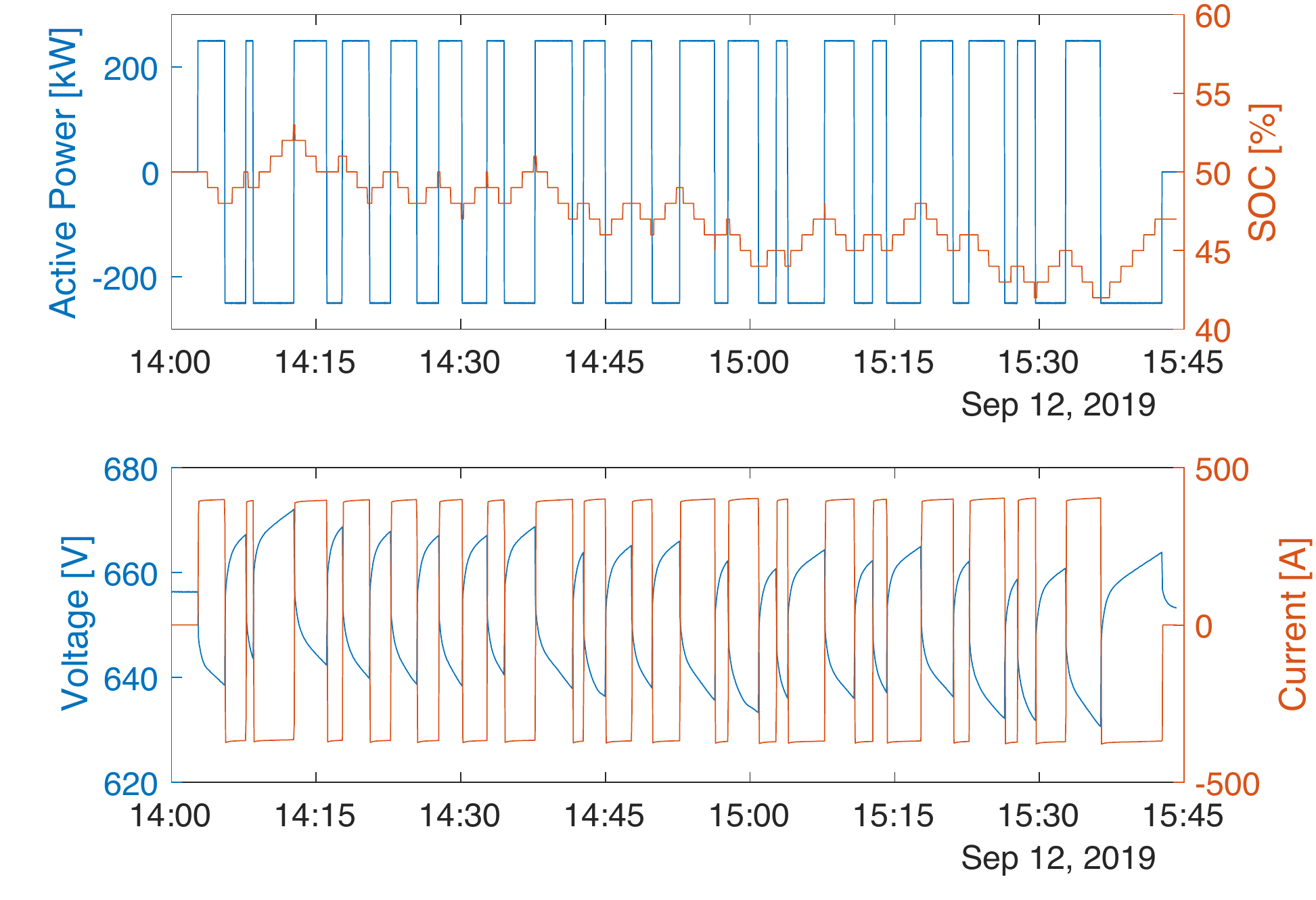}
		\caption{PRBS active power reference and measured SOC, $v^{DC}$ and $i^{DC}$ for the SOC range of 34-66\% in case of 7 strings in operation.}
		\label{fig:TTC_model_test}
\end{figure}

A dedicated experimental investigation allowed the estimation of the equivalent TTC circuit parameters via a grey-box modeling-based approach, in line with the analogue estimation activity proposed in~\cite{Emil_ISGT} for the same BESS in case of full available storage capacity (9 strings). 
The model identification tests are based on pseudo-random binary sequence (PRBS), i.e., a two levels square wave with on-off periods of normally distributed random durations, capable of exciting a wide range of system dynamics. 
Fig. \ref{fig:TTC_model_test} shows the binary power set-points, the SOC, $v_{dc}$ and $i_{dc}$.
Since the TTC model parameters depend on the BESS SOC, the test has been repeated for different SOC ranges.
The obtained TTC model parameters are in Table~\ref{ttc}. 

\begin{table}[!ht]
\centering
		\renewcommand{\arraystretch}{1}
\caption{Estimeted BESS Parameters for 7 strings for different SOC ranges}
\label{ttc}
\begin{tabular}{|l|c|c|c|l}
\cline{1-4}
                & SOC=0-33\% & SOC=34-66\% & SOC=67-100\% &  \\ \cline{1-4}
$a$                       & 607.2           & 607.1            & 590.0             &  \\
$b$                        & 190.8           & 113.9            & 188.9             &  \\
$R_{s}$ [$\Omega$]                       & 0.0221          & 0.0165           & 0.0155            &  \\
$R_{1}$ [$\Omega$]                       & 0.0131          & 0.0120           & 0.0109            &  \\
$C_{1}$ [$F$]                        & 1511            & 1844             & 1917              &  \\
$R_{2}$ [$\Omega$]                       & 5.26E-05        & 2.24E-05         & 2.55E-04          &  \\
$C_{2}$ [$F$]                        & 1.00E+06        & 1.00E+06         & 1.00E+06          &  \\
$R_{3}$ [$\Omega$]                       & 5.10E-06        & 6.50E-07         & 1.55E-05          &  \\
$C_{3}$ [$F$]                        & 1.00E+07        & 1.00E+07         & 1.00E+07          &  \\ \cline{1-4}
\end{tabular}
\end{table}

With reference to \eqref{eq:droop_set}, in order to properly set the initial values of the droop constants $\alpha_{0}$ and $\beta_{0}$, the maximum deliverable active and reactive powers $P^{max}$ and $Q^{max}$ have been used along with the calculated maximum deviation of the input variables of the controller, i.e., $\Delta^{max} f$ and $\Delta^{max} v^{AC}$. 

Historical measurements acquired by the synchrophasor network on the EPFL MV network are used for this purpose, whose P-class phasor measurement units (PMUs) allowed the acquisition of data with a timestamp of 20 ms~\cite{Pignati}.
The values of $\Delta^{max} f$ and $\Delta^{max} v^{AC}$ have been obtained by approximating their distribution with normal distribution functions and by considering a relevant multiplication factor for the standard deviations $\sigma$. 
On the one hand, the maximum deviations of $\pm$3.3$\sigma_{f}$ was considered for the system frequency measurements, meaning that the thresholds $\mu_{f}$$\pm$3.3$\sigma_{f}$ are statistically exceeded only 0.1$\%$  of the times, being $\mu_{f}$ the average value of the frequency dataset, equal to 50.000 Hz. 
This rather strict assumption is motivated by the requirement from the Swiss TSO grid code on the quality of the supply of primary frequency control power, which sets a maximum tolerable time of 0.1$\%$ of the tender period for which the regulating power cannot be delivered without running into penalties~\cite{Swiss_PFC_codes}.
On the other hand, since less strict requirements regulate the quality of the supply of local voltage control, smaller maximum deviations can be considered: the calculated thresholds for the activation of the maximum reactive power capacity are $\mu_{V}$$\pm$1$\sigma_{V}$, where $\mu_{V}$ is the average value of the AC phase-to-phase voltage dataset, equal to 21.192 kV.
Since the obtained $\mu_{V}$ differs from the nominal value of 21 kV, it was decided to consider $\mu_{V}$ as reference for the calculation of $\Delta^{max} v^{AC}$ in \eqref{eq:droop_p_q}.

Given the considered historical dataset, $\Delta^{max} f$ = $\pm$3.3$\sigma_{f}$ = $\pm$58.8 mHz and $\Delta^{max} v^{AC}$ = $\pm$1$\sigma_{V}$ = $\pm$0.0672 kV. 
The calculated $\Delta^{max} f$ and $\Delta^{max} v^{AC}$ enable the computation of the initial droops $\alpha_{0}$ and $\beta_{0}$ for different BESS configurations considering the number of available strings, i.e., the shrink factors $C_{shrink}$, as shown in Table~\ref{Droops}. 

\begin{table}[!ht]
\centering
		\renewcommand{\arraystretch}{1}
\caption{Calculated $\alpha_{0}$ and $\beta_{0}$ for different shrink factors $C_{shrink}$}
\label{Droops}
\begin{tabular}{|l|c|c|c|l}
\cline{1-3}
$C_{shrink}$       & $\alpha_{0} [kW/Hz]$ & $\beta_{0} [kvar/V]$           \\ \cline{1-3}
$1$                       & 11575           & 10.78 \\
$8/9$                        & 10289           & 9.58 \\
$\textbf{7/9}$                       & \textbf{9003}           & \textbf{8.39} \\
$6/9$                        & 7717           & 7.19 \\
$5/9$                       & 6430           & 5.99 \\
$4/9$                        & 5144           & 4.79 \\
$3/9$                       & 3858           & 3.59  \\
$2/9$                        & 2572           & 2.40 \\
$1/9$                        & 1286           & 1.20  \\ \cline{1-3}
\end{tabular}
\end{table}

\vspace{-0.1 cm}

\begin{figure}[ht]
		\centering
    	\includegraphics[width=0.95\linewidth]{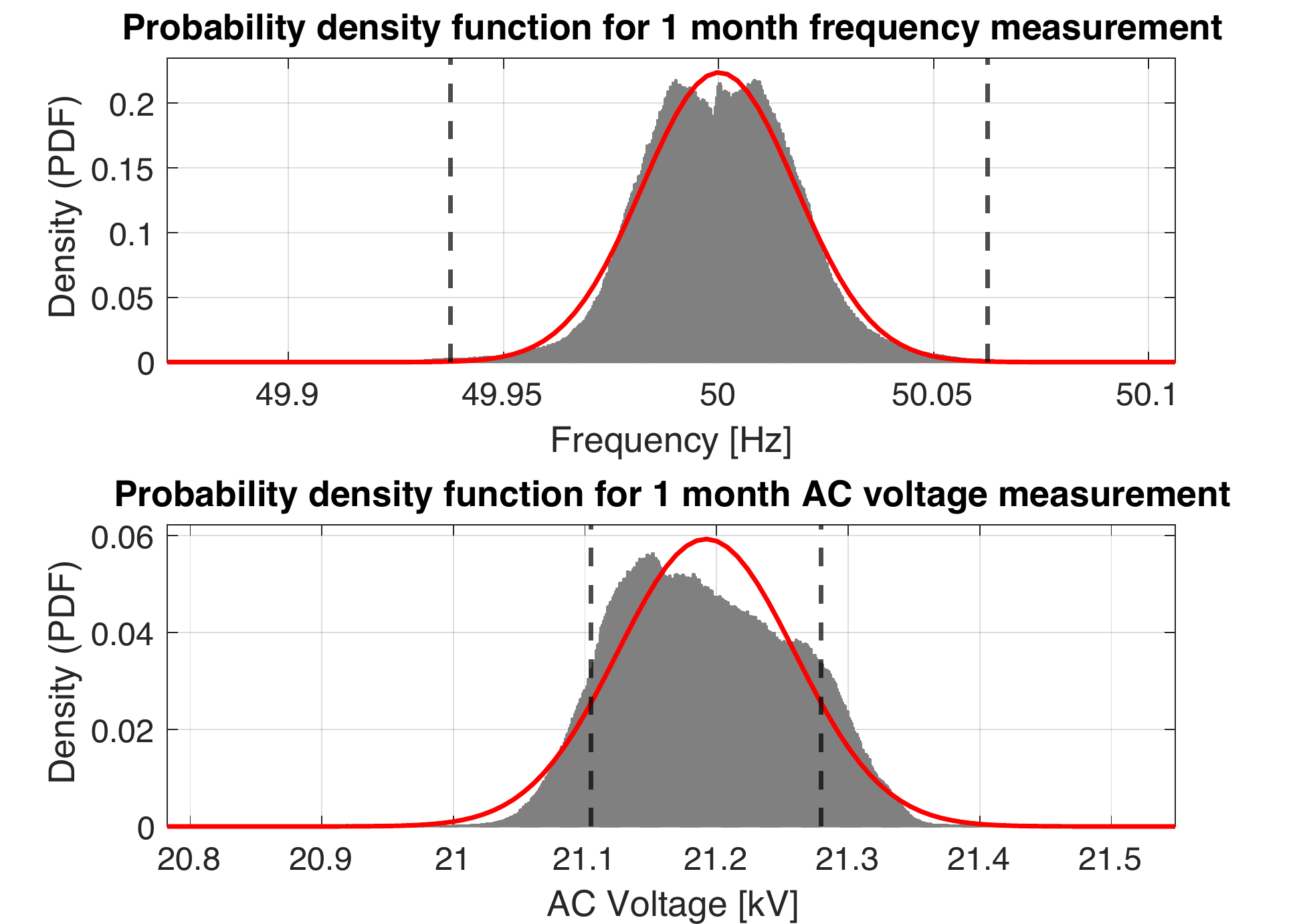}
		\caption{1 month historical data of frequency and phase-to-phase voltage at the BESS PCC at 21 kV, acquired via PMUs installed at the EPFL MV network. The dashed lines represent the limits of $\mu_{f}$$\pm$3.3$\sigma_{f}$ and $\mu_{V}$ $\pm$1$\sigma_{V}$ for frequency and voltage measurements, respectively.}
		\label{fig:PMU_hist_meas}
\end{figure}

\section{Experimental Investigation}
A number of scenarios have been investigated considering different combinations of initial droops $\alpha_{0}$ and $\beta_{0}$.
Table~\ref{Scenarios} reports the overview of the analysed cases. Each test has been carried out for a 5-minute time window, and the real-time BESS battery and AC grid statuses have been monitored and processed in order to compute the optimal P and Q set-points as described in Section II. 
The same priority has been given to the provision of P and Q by setting $\lambda_{P}$=$\lambda_{Q}$=1 in the implementation of \eqref{eq:opt_pq}.  
A time granularity of 1 second has been used for data acquisition and optimal set-point computation, meaning that at each second a new operating point within the corresponding feasible PQ region is sent to the BESS converter controller.
The choice of 1-second response is considered as a realistic assumption in BESS applications as indicated, for instance, by the newly-released grid code by the Danish TSO Energinet.dk \cite{DK_BESS_code}. However, the Authors are aware that in low-inertia power systems rapid (i.e., sub-second) frequency variations are more likely to be experienced \cite{AUS_Blackout_Report}, meaning that even faster response from control providers may be needed.

\begin{table}[!ht]
\centering
		\renewcommand{\arraystretch}{1}
\caption{Calculated $\alpha_{0}$ and $\beta_{0}$ for different shrink factors $C_{shrink}$}
\label{Scenarios}
\begin{tabular}{|l|c|c|c|l}
\cline{1-3}
$Scenario$       & $\alpha_{0}$ & $\beta_{0}$           \\ \cline{1-3}
$\#1$                       & 9003 [kW/Hz]           & 8.39 [kvar/V]
 \\
$\#2$                        & 9905 [kW/Hz]           & 8.39 [kvar/V]
 \\
$\#3$                       & 19810 [kW/Hz]           & 8.39 [kvar/V]
 \\
$\#4$                        & 29715 [kW/Hz]           & 12.57 [kvar/V]
 \\ \cline{1-3}
\end{tabular}
\end{table}

\begin{figure}[ht]
		\centering
    	\includegraphics[width=0.85\linewidth]{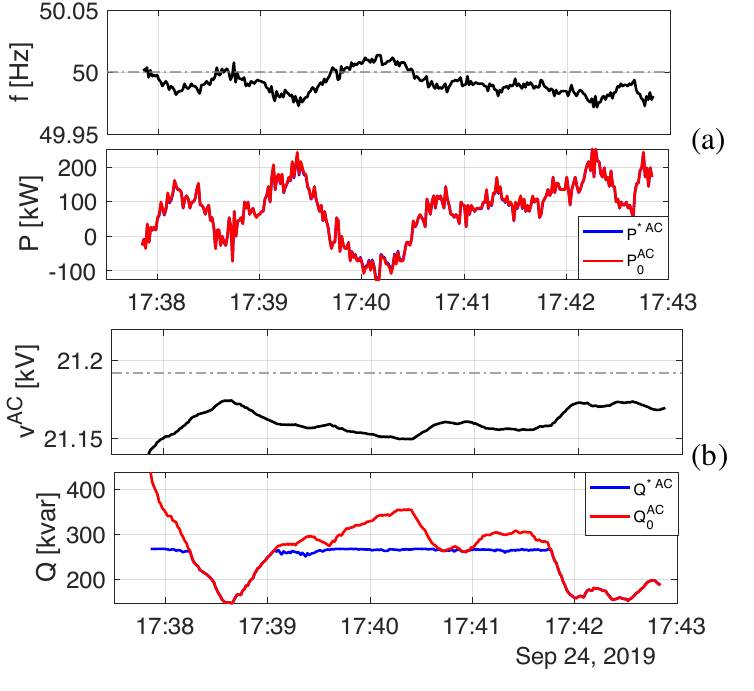}
		\caption{$Scenario \#1$ Results. (a): PFC; (b): local voltage control.}
		\label{fig:Scenario_1_CTRL}
\end{figure}

Fig. \ref{fig:Scenario_1_CTRL} shows results for $Scenario \#1$, for which $\alpha_{0}$ and $\beta_{0}$ are calculated as in Section III. 
The top subplots of \textit{(a)} and \textit{(b)} report the measured AC grid frequency and the mean value of the three phase-to-phase voltages at the MV connection point, with the respective reference values used for the calculation of $\Delta f_{t}$ and $\Delta v^{AC}_{t}$ as in \eqref{eq:droop_p_q}.
The computed P-Q set-point calculated implementing the standard droop control equation in \eqref{eq:droop_p_q} are reported in red in the bottom subplots.
Additionally, the actual set-points computed as result of the optimization problem are shown with the blue lines. 
Note that, as already stated in Section II, for the selection of the appropriate converter PQ capability curve at each time-step, the AC voltage measured at the 21 kV busbar is scaled down to the LV side voltage level of the BESS step-up transformer using the associated transformation ratio.
Firstly, it can be seen that for frequency measurements larger than 50 Hz, the BESS behaves as a load: the sign of the exchanged active power is negative, meaning that the BESS is charging. 
Symmetrically, when the frequency is below the 50 Hz, the BESS discharges by injecting active power with positive sign into the grid.
Similar considerations are valid for the local voltage control. 
In general, for $\Delta v^{AC} > 0$, i.e., in case of over-voltages, negative reactive power is provided by the BESS, meaning that the BESS behaves as an inductor. 
On the contrary, for $\Delta v^{AC} < 0$, i.e., in case of under-voltages, capacitive reactive power is provided, as in the case of the whole time-window for the test of $Scenario \#1$.
Secondly, it can be noticed that the desired primary frequency support is fully achieved since the expected active power is provided at any moment of the considered time window. 
By contrast, the relatively large value of the initial droop $\beta_{0}$ and the measured deviations of the AC voltage from the reference value, caused a mismatch between the expected and the provided voltage control service for more than half of the time of the test. 
In fact, in these cases the desired Q set-point would have been out of the feasible region of the employed hardware, hence the proposed optimal control approach moved it to the edge of the corresponding PQ capability curve. 

\begin{figure}[ht]
		\centering
    	\includegraphics[width=0.85\linewidth]{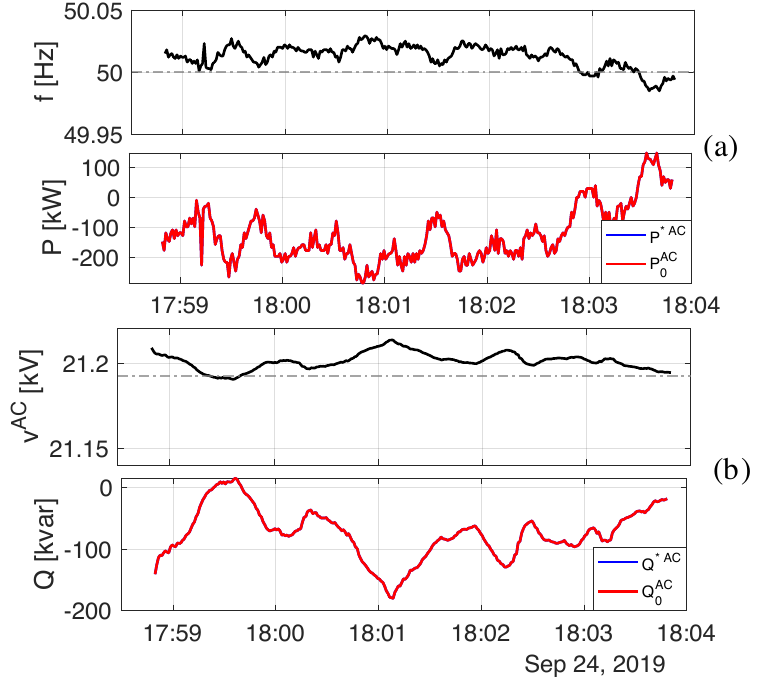}
		\caption{$Scenario \#2$ Results. (a): PFC; (b): local voltage control.}
		\label{fig:Scenario_2_CTRL}
\end{figure}

The test for $Scenario \#2$ presented in Fig. \ref{fig:Scenario_2_CTRL} shows a case when the local voltage control via reactive power is achieved continuously, although the implemented initial droop $\beta_{0}$ is the same as in $Scenario \#1$.
Also the primary frequency control action is performed continuously, responding as desired to the measured frequency signal for the whole duration of the test. 
In this case a larger initial droop $\alpha_{0}$ was implemented, namely a value calculated considering $\pm$3$\sigma_{f}$ as the maximum frequency deviation, i.e., with a confidence interval of 99.7$\%$. 

\begin{figure}[ht]
		\centering
    	\includegraphics[width=0.85\linewidth]{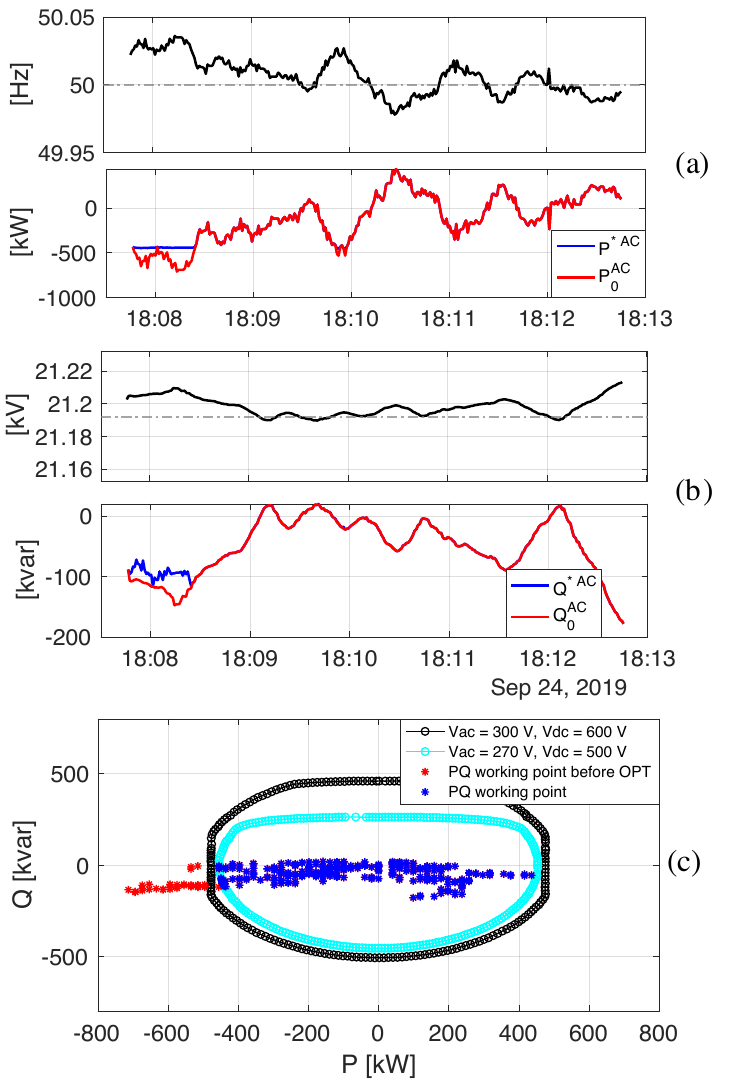}
		\caption{$Scenario \#3$ Results. (a): PFC; (b): local voltage control; (c): PQ set-points before and after the proposed optimization algorithm.}
		\label{fig:Scenario_3_CTRL}
\end{figure}

In $Scenario \#3$ an even larger value of $\alpha_{0}$ was used, corresponding to $\pm$1.5$\sigma_{f}$ as the maximum frequency deviation, i.e., with a confidence interval of 86.6$\%$.
In this test case, the solution of the optimization problem enabled the BESS to operate also when the calculated P set-points falls outside the feasible region of the considered PQ capability curve.  
In fact, the values at the edge of the feasible region were set, meaning that the frequency service was not performing as desired, although the maximum power was still provided to partially support the grid. 
From Fig. \ref{fig:Scenario_3_CTRL}, it can be seen that this happens in the first 38 seconds of the test and for a shorter period of time also around the mid point.
Fig. \ref{fig:Scenario_3_CTRL}-\textit{(c)} maps the operating points before and after the implementation of the proposed optimal set-point calculation. 
It can be noticed that, thanks to the proposed method, the points falling outside the feasible region have been retrieved to the edge of the light blue converter feasible region, thus assuring the continuity of the delivery of the two grid services.
Furthermore, it is of paramount importance to note that without the proposed optimal controller the too high value of the computed P set-point would have made the BESS converter either trip or go to 0 kW for safety reasons.
Under these circumstances, the expected service would have been fully undelivered, enhancing the probability of reaching the 0.1$\%$ threshold imposed by the Swiss TSO for undelivered regulating power when providing primary frequency control.
It is in fact relevant to quantify the amount of regulating energy actually delivered during the regulation session and to compare it with the energy that would have been delivered without optimization and in the ideal case of un-constrained BESS power converter.
So, the quantification - as for PFC provision - of the concrete effects of the proposed controller with control time granularity $\Delta t$ during a control session of duration $T \Delta t$ is done as described in equations \eqref{eq:E_exp}-\eqref{eq:E_NONopt}.
They define the expected energy $E_{exp}$, the actual delivered energy with the optimal control $E^{*}$, and the energy that would have been delivered without the proposed optimal approach $E_{0}$, respectively. 
Such quantification is included in Table~\ref{Results_Energy}.


\begin{align}
    E_{exp} = \sum_{i=1}^{T} \Delta t \left | \alpha_0 \Delta f_i \right | \label{eq:E_exp}
\end{align}

\begin{align}
    E^{*} = \sum_{i=1}^{T} \Delta t \left | P^{*AC}_i \right | \label{eq:E_opt}
\end{align}

\begin{align}
    E_{0} = \sum_{i=1}^{T} \Delta t \left | P^{AC}_{0, i} \right | \label{eq:E_NONopt}
\end{align}


\begin{figure}[ht]
		\centering
    	\includegraphics[width=0.88\linewidth]{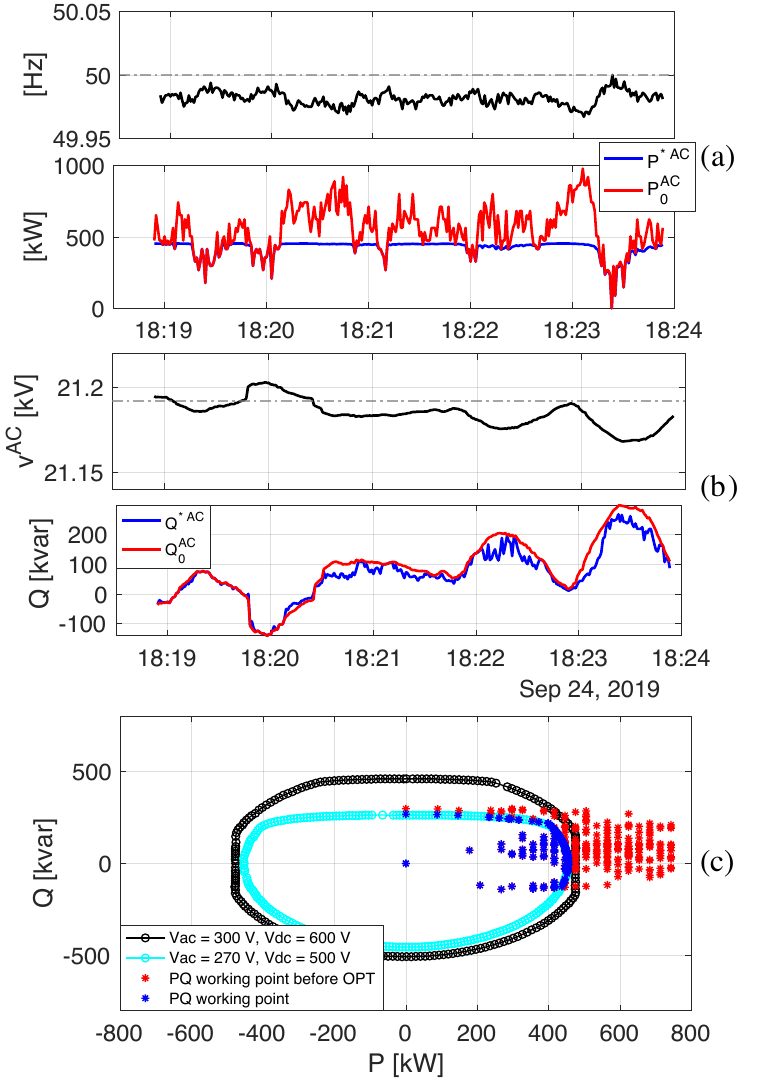}
		\caption{$Scenario \#4$ Results. (a): PFC; (b): local voltage control; (c): PQ set-points before and after the proposed optimization algorithm.}
		\label{fig:Scenario_4_CTRL}
\end{figure}

Finally, $Scenario \#4$ is analysed to assess the situation in case of a very large initial droops $\alpha_{0}$ and $\beta_{0}$, corresponding to maximum deviations of $\pm$1.5$\sigma_{f}$ and $\pm$0.75$\sigma_{V}$.
Although the very low measured voltage deviations made the computed Q set-point be inside the feasible PQ region all the time, the same is not valid for P.
In fact, Fig. \ref{fig:Scenario_4_CTRL} shows that for almost the whole duration of the test, the P set on the converter is at the edge of the selected feasibility curve, meaning that the primary frequency grid service is not fully delivered. 
The mapping of the PQ set-points before and after the solution of the proposed optimization problem is shown in Fig. \ref{fig:Scenario_4_CTRL}-\textit{(c)}. As for $Scenario \#3$, the continuity of the delivery of the two grid services is possible thanks to the projection of the initially-calculated set-points to the edge of the light blue converter feasible region.
The quantification of the effectiveness of the optimal controller in terms of expected and delivered regulating energy is reported in Table~\ref{Results_Energy}.

\begin{table}[!ht]
\centering
		\renewcommand{\arraystretch}{1}
\caption{Expected and delivered energy for PFC with and without the proposed optimization algorithm}
\label{Results_Energy}
\begin{tabular}{|l|c|c|c|c|l}
\cline{1-4}
$Scenario$       & $E_{exp}$ & \quantities{$E^{*}$} & \quantities{$E_{0}$}           \\ \cline{1-4}
$\#1$                       & 8.3 [kWh]           & 8.3 [kWh]   & 8.3 [kWh]
 \\ \hline
$\#2$                       & 10.5 [kWh]           & 10.5 [kWh]   & 10.5 [kWh]
 \\ \hline
$\#3$                       & 20.0 [kWh]           & \quantities{18.4 [kWh] \\ (91.90$\%$)}  & \quantities{13.7 [kWh] \\ (68.47$\%$)}
 \\ \hline
$\#4$                        & 45.7 [kWh]           & \quantities{35.0 [kWh] \\ (76.49$\%$)}  & \quantities{9.1 [kWh] \\ (19.91$\%$)}
 \\ \cline{1-4}
\end{tabular}
\end{table}

\section{Conclusions}
The work presented a BESS control framework for optimal provision of concurrent power system services. 
In particular, primary frequency and local voltage control are achieved via the modulation of active and reactive power set-points, respectively, exploiting the flexibility given by the 4 quadrant power converter. 
The proposed algorithm considers the working conditions of the AC utility grid as well as the battery DC voltage as a function of the current trajectory using the battery TTC model, in order to select the suitable converter capability curve, which is not unique for all the possible operating conditions, hence optimizing the provision of grid services. 
A computationally-efficient algorithm was proposed to solve the formulated optimal power set-points calculation problem.

A set of experimental tests on a commercial utility-scale 720 kVA/560 kWh BESS showed the capability of the controller to enable PFC and local voltage control not only by charging or discharging the battery, but also by means of reactive power exchange, namely behaving as inductor or capacitor in case of over- or under-voltages, respectively. 
When in case of large initial droop constants or large frequency/voltage deviations the PQ feasible region is passed, the proposed controller enabled the operation at the edge of the selected PQ capability curve, dramatically reducing the amount of accumulated non-delivered regulating power during the control session.   
Hence, the paper highlighted the importance of accurately modelling the employed hardware in order to enable an optimal grid service provision even under non-nominal BESS conditions (e.g., reduced available number of strings) as well as under commercial hardware embedded technical limitations (e.g., variable capability curves of the power converter).

Future works include the extension of the complexity of the model by considering the power conversion efficiency as a function of the exchanged AC active power, and a series of experimental tests to map more systematically all the possible capability curves for a wider range of combinations of battery DC voltage and grid voltage conditions.
Further, investigations on BESS control logics as voltage source in combination with the provision of ancillary services are of interest.  

\bibliographystyle{IEEEtran}
\bibliography{biblio}{}
\end{document}